\documentclass[]{aa}  

\usepackage{graphicx}
\usepackage{txfonts}
\usepackage{lipsum}
\usepackage{subcaption}         
\usepackage{lscape}             
\usepackage{placeins}           

\usepackage{booktabs}
\usepackage{amsmath, amssymb}
\usepackage{graphicx}
\usepackage{booktabs}
\usepackage{natbib}
\usepackage{hyperref}
\usepackage{lineno}
\usepackage{xcolor}

\titlerunning{Bistability and Stochastic Resonance in Binary Planetary Systems}
\begin{document}

   \title{Resonant Hamiltonian Dynamics in the CR3BP: \\
Bistability and Stochastic Resonance in Binary Planetary Systems}


%

   \author{R. Capuzzo-Dolcetta\inst{1,2,3,4}\fnmsep\thanks{Corresponding author: \\roberto.capuzzodolcetta@fondazione.uniroma1.it}}

   \institute{
   Dipartimento di Fisica, Sapienza, Universit\'a di Roma, P.le A.Moro 2, I-00185, Roma, Italy
\and INFN, Sezione di Roma, Sapienza, Universit\'a di Roma, P.le A.Moro 2, I-00185, Roma, Italy
             \and INAF, Observatory of Roma at Monteporzio Catone, via Frascati, 33, I-00078, Monteporzio Catone, Italy 
             \and
             Centro Ricerche Enrico Fermi,
             Piazza del Viminale 1,
             I-00184 Roma, Italy
}
   \date{Received March XX, 20XX}

 

\abstract
{The Circular Restricted Three-Body Problem (CR3BP) provides a fundamental framework for understanding resonant dynamics in binary star systems.} 
{We develop a unified Hamiltonian formulation for mean-motion resonances that encompasses both circumstellar (S-type) and circumbinary (P-type) planetary orbits within the CR3BP. Unlike the Solar System case where the perturbing body is a planet of negligible mass, here the perturber (a stellar companion) has a non-negligible, finite mass – a crucial difference that we fully incorporate.}
{Starting from the full Hamiltonian in each configuration, we perform canonical transformations to resonant action-angle variables and derive reduced one-degree-of-freedom Hamiltonians through systematic averaging over the fast orbital motion.  
Leading‑order scaling laws for the Fourier coefficients of the resonant perturbation are obtained, revealing their dependence on the binary mass ratio and the planet’s orbital distance.
}
{The resulting effective potential is shown to exhibit bistability under the well-defined condition $|\epsilon_2/\epsilon_1| > 1/4$, where $\epsilon_1$ and $\epsilon_2$ are the amplitudes of the first two resonant harmonics. This bistability creates the essential dynamical setting for stochastic resonance. Scaling laws for the Fourier coefficients are derived for both S-type and P-type configurations. Estimates for known binary-planet systems (including Kepler-16b, Kepler-34b, and Gamma Cephei Ab) show that while currently observed systems lie below the bistability threshold, the theory predicts that extreme configurations ($a/a_b \lesssim 1.5$ for P-type, almost equal mass binary) could host bistable resonances accessible to future observations.} 
{This work provides a natural Hamiltonian framework for studying stochastic resonance in binary planetary systems, bridging analytical celestial mechanics and the nonlinear dynamics of exoplanetary systems subject to realistic perturbations.}

\keywords{celestial mechanics — planets and satellites: dynamical evolution and stability — binaries: general — methods: analytical}

\maketitle

\nolinenumbers

\section{Introduction}
\label{sec:intro}
\begin{figure}[ht]
\centering
   \includegraphics[width=0.8\linewidth]{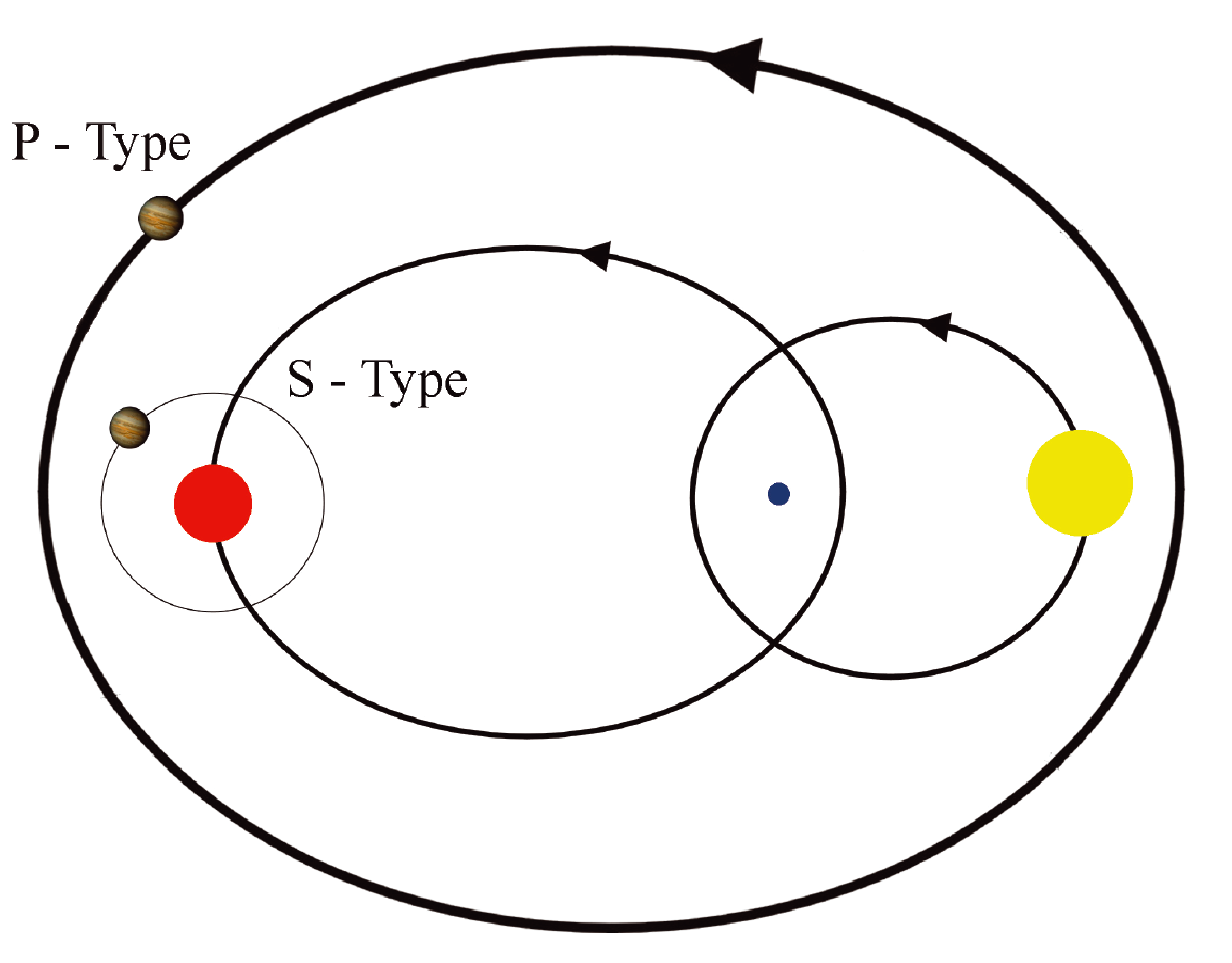}
\caption{Schematic geometry of S-type (circumprimary) and P-type (circumbinary) planetary orbits. The binary stars A (red) and B (yellow) orbit their common barycenter (black dot). The planet (brown) on the S-type orbit revolves around A; the P-type orbit corresponds to its revolution around the binary barycenter.}
\label{fig:1}
\end{figure}

The discovery of planets in binary and multiple star systems has dramatically expanded the landscape of exoplanetary science \citep{doyle2011,schwarz2016,schwarz2018,schwarz2021}. Two main orbital architectures are observed: \emph{S-type} (circumstellar) planets, which orbit one member of the binary, and \emph{P-type} (circumbinary) planets, which orbit both stars (see Fig. \ref{fig:1}).
Reliable recent estimates \citep{nasaexoplanetarchive2024,martin2019} report of about $50-55$ observed planets on P-type orbits and about 10 times more ($450-500$) on S-type orbits. Before 2010, almost all planets around binaries were found on S-type orbits, while from 2011 to 2018 the Kepler satellite began discovering circumbinary planets. The following TESS satellite (launched in 2018) increased significantly the number of discoveries of planets revolving on P-orbits \citep{kostov2020}. For a recent statistical analysis of binary-planet populations see \citet{society2023} for a statistical update on binary planet populations.

The Circular Restricted Three-Body Problem (CR3BP) provides the natural dynamical framework for analyzing these systems. In this work, we develop a unified Hamiltonian formulation for mean-motion resonances within the CR3BP that encompasses both S-type and P-type configurations.

The dynamical environment in such S and P systems is rich, with mean-motion resonances driven by the gravitational interplay between the planet and the binary pair. Resonant dynamics can govern orbital stability, migration, and long-term evolution, making their analytical description a priority for understanding observed systems and predicting new ones.

While resonant phenomena in the CR3BP have been studied extensively \citep{murray1999, winter1997}, the role of {stochastic effects} in resonant transitions remains comparatively unexplored. In particular, {stochastic resonance} (SR)---a nonlinear mechanism where noise amplifies a weak periodic signal in a bistable system---was first identified by \cite{benzi1981} and applied, first,  to climate dynamics \citep{benzi1982} and later established as a generic phenomenon in noisy nonlinear systems \citep{gammaitoni1998}. Its mathematical formulation in perturbed Hamiltonian systems was further developed by \citet{freitas2005}, among others. In celestial mechanics, SR has been invoked to explain noise‑induced transitions in asteroid resonances \citep{morbidelli2002} and rotational dynamics of small bodies \citep{shevchenko2011}, but it has never been systematically applied to binary planetary systems. 

SR requires three ingredients: (i) a bistable potential, (ii) weak periodic forcing, and (iii) a source of noise. Binary systems naturally provide all three: resonant potentials can become bistable through higher‑order harmonics; slow secular perturbations or migration supply periodic forcing; and stochastic perturbations arise from disk remnants, stellar activity, or chaotic interactions.

The goal of this paper is to derive a unified \emph{resonant Hamiltonian framework} for both S-type and P-type orbits in the CR3BP, and to show how this framework naturally leads to a bistable potential capable of supporting stochastic resonance. 



To the best of our knowledge, this work provides the first unified Hamiltonian formulation for mean-motion resonances in both S-type and P-type binary planetary systems that is systematically reduced to a one-degree-of-freedom resonant model. The derivation yields closed-form expressions for the Fourier coefficients directly linked to observable parameters, and shows that the resulting resonant potential becomes bistable under well-defined, observationally testable conditions — the essential dynamical ingredient for stochastic resonance. This formalism thus bridges classical resonant perturbation theory and the emerging study of stochastic phenomena in exoplanetary dynamics.

\section{CR3BP Setup and Hamiltonian Separation}
\label{sec:setup}
We adopt the standard CR3BP normalization: total mass $M_A+ M_B = 1$, binary separation $a_b = 1$, gravitational constant $G = 1$, 
from which it follows that the binary angular velocity $\omega_b=1$. We consider star $A$ as primary, and $M_A>M_B$; consequently, we define the mass ratio as  $\nu = M_B/(M_A+M_B)$, so that  $M_A = 1 - \nu$ and $M_B = \nu$.

\subsection{S-type Orbits}
\label{S-type Orbits}
For S-type orbits (planet bound to $M_A$), the full Hamiltonian in the barycentric rotating frame is

\begin{equation}
\label{eq:1}
H^S = \frac{1}{2}(p_x^2 + p_y^2) + (y p_x - x p_y) - \frac{1}{2}(x^2 + y^2) - \frac{1-\nu}{r_A} - \frac{\nu}{r_B},
\end{equation}
where  $x,y$ are the Cartesian coordinates of the planet, $p_x,p_y$ are the canonical momenta, and 
\begin{equation}
\label{eq:2}
r_A = \sqrt{(x+\nu)^2 + y^2}, \quad r_B = \sqrt{(x-1+\nu)^2 + y^2},
\end{equation}

the distances to $A$ and $B$, respectively.
We separate $H^S = H_0^S + H_1^S$ with
\begin{align}
H_0^S &= \frac{1}{2}(p_x^2 + p_y^2) + (y p_x - x p_y) - \frac{1}{2}(x^2 + y^2)- \frac{1-\nu}{r_1}, \label{eq:3} \\
H_1^S &= -\frac{\nu}{r_2}. \label{eq:4}
\end{align}
Here $H_0^S$ describes Keplerian motion around $M_1$ in the rotating frame, and $H_1^S$ is the direct perturbation from $M_2$.

\subsection{P-type Orbits}
For P-type orbits (planet orbiting the barycenter), the full Hamiltonian is, obviously, the same as before
\begin{equation}
\label{eq:5}
H^P = \frac{1}{2}(p_x^2 + p_y^2) + (y p_x - x p_y) - \frac{1}{2}(x^2 + y^2) - \frac{1-\nu}{r_A} - \frac{\nu}{r_B},
\end{equation}
but the natural unperturbed motion is Keplerian around the total mass at the barycenter. Thus we set
\begin{align}
H_0^P &= \frac{1}{2}(p_x^2 + p_y^2) - \frac{1}{r}, \label{eq:6} \\
H_1^P &= (y p_x - x p_y) - \frac{1}{2}(x^2 + y^2) - \left[ \frac{1-\nu}{r_A} + \frac{\nu}{r_B} - \frac{1}{r} \right], \label{eq:7}
\end{align}
where $r = \sqrt{x^2 + y^2}$. For $r > 1$, the binary potential (first two addends in the squared bracket above) can be expanded in multipoles (up to $n=2$):

We emphasize that our partition of the Hamiltonian for P-type orbits places the Coriolis and centrifugal terms in the perturbative part \(H_1^P\), rather than in \(H_0^P\). An alternative separation that includes these terms in \(H_0^P\) would lead to the so-called rotating Kepler problem. Although that problem is integrable, its solution is less elementary and does not admit simple, closed-form action-angle variables of the Delaunay type. By keeping \(H_0^P\) as the standard inertial Kepler problem, we retain the full power of the classical Delaunay transformation and the averaging method, which are the most direct tools to derive the reduced resonant Hamiltonian and the bistability condition. This choice is therefore a matter of methodological convenience and clarity, not of physical necessity.

\begin{equation}
\label{eq:8}
-\left[\frac{1-\nu}{r_A} + \frac{\nu}{r_B}\right] = -\frac{1}{r} -  \frac{\nu(1-\nu)}{2r^3}(3\cos^2\theta - 1) + \mathcal{O}(r^{-5}),
\end{equation}

simplifying $H_1^P$ for analytical treatment.

The unperturbed Hamiltonian $ H_0^P $ in Eq.~\ref{eq:6} describes Keplerian motion around a point mass of unit total mass located at the barycenter. This choice corresponds to taking an \emph{inertial frame} whose origin coincides with the barycenter and in which the binary stars orbit each other. In such an inertial frame, the gravitational potential experienced by the planet is, to lowest order, simply that of a single central mass $ M_A+M_B = 1 $.

When we move to the \emph{rotating frame} that co‑rotates with the binary (the frame in which the stars are fixed at $ (-\nu,0) $ and $ (1-\nu,0) $), the Hamiltonian acquires additional terms: the Coriolis term $ (y p_x - x p_y) $ and the centrifugal potential $ -(x^2+y^2)/2$. Because these terms arise solely from the change of reference frame and are absent in the inertial Kepler problem, they are naturally grouped into the perturbative part $ H_1^P $, together with the higher‑order multipolar corrections of the binary potential (the term in square brackets in Eq.~\ref{eq:7}).

This separation contrasts with the S‑type case (Section~\ref{S-type Orbits}), where the unperturbed motion is defined directly in the rotating frame around the primary star; there, the Coriolis and centrifugal terms are already included in $ H_0^S $. The different treatment reflects the natural reference frame for each configuration: for S‑type orbits the primary star is the natural origin, while for P‑type orbits the barycenter of the binary is the natural origin of an inertial description.

Before proceeding, it is worth noting the correspondence with the well-known problem of interior and exterior mean-motion resonances in the Solar System. The S-type configuration (planet around the primary, perturbed by the secondary star) is dynamically analogous to an \textit{interior} resonance, where the perturber lies outside the planet's orbit. Conversely, the P-type configuration (planet around the barycenter, perturbed by the binary) resembles an \textit{exterior} resonance, where the perturber lies inside the planet's orbit. However, a key difference is that in binary star systems the secondary star has mass ratio \(\nu\) that can be of order unity (up to \(0.5\)), whereas in the Solar System the perturbing planet satisfies \(\nu \ll 1\). This difference dramatically affects the amplitude of the resonant Fourier coefficients. In particular, the ratio \(|\epsilon_2/\epsilon_1|\), which controls the bistability condition derived in Section~6, can become large enough even at low eccentricity in the binary case, allowing bistability for resonances that remain monostable in the planetary case. The following sections quantify this effect.

\section{Transformation to Action-Angle Variables}

In the following, we define Delaunay variables separately for S‑type and P‑type orbits. To avoid confusion between the two configurations, we attach a subscript \(S\) or \(P\) to the canonical variables \(\Lambda, \lambda, L, \varpi\). The physical orbital elements \(a, e, \mathcal{M}, \omega\) are the same geometric quantities in both cases and carry no subscript. For S‑type orbits, the variables are defined in the rotating frame (primary fixed); for P‑type orbits, they are defined in the inertial barycentric frame. The two sets are related by a time‑dependent rotation: \(\lambda_P = \lambda_S + \lambda_b\) and \(\varpi_P = \varpi_S + \lambda_b\), where \(\lambda_b = t + \mathrm{const}\) is the mean longitude of the binary. This relation follows from \(\varpi_{\text{rot}} = \varpi - \lambda_{\text{perturber}}\) as suggested by the referee. For resonance analysis we work directly in the frame natural to each configuration, and the two sets of variables are never mixed.

\subsection{S-type Orbits}
\label{subsec:S-Delaunay}

For the unperturbed Keplerian motion around the primary $ M_A $ in the rotating frame, we introduce the classical Delaunay variables. They are defined with respect to the Kepler problem with central mass $ M_A = 1-\nu $. In terms of the orbital elements:

\begin{align}
    \Lambda_S &= \sqrt{(1-\nu) a}, \quad & \text{(radial action)} \label{eq:9}\\
    \lambda_S&= \cal{M} + \varpi^{(S)}, \quad & \text{(mean longitude)} \label{eq:10}\\
    L_S &= \Lambda_S \sqrt{1-e^{2}}, \quad & \text{(magnitude of angular momentum)} \label{eq:11}\\
    \varpi_S &= \omega, \quad & \text{(longitude of pericenter.)} \label{eq:12}
\end{align}

Here $ a $ is the semimajor axis relative to the primary, $ e $ the planet orbit eccentricity, $ \cal{M} $ the mean anomaly, and $ \omega $ the argument of pericenter measured from the rotating $ x $-axis. 
In the coplanar problem, the orbital plane coincides with the binary plane; therefore the longitude of the ascending node $ \Omega $ is undefined (or can be set to zero), and the longitude of pericenter $ \varpi $ simply equals the argument of pericenter $ \omega $.

The unperturbed Hamiltonian in Eq. \ref{eq:3} becomes simply

\begin{equation}
\label{eq:13}
    H_0^{S} = -\frac{(1-\nu)^{2}}{2\Lambda_S^{2}}.
\end{equation}

The perturbative part $ H_1^S $ (Eq. \ref{eq:4}), when expressed in the same Delaunay variables, remains a function of both actions and angles:

\begin{equation}
\label{eq:14}
H_1^S = H_1^S(\Lambda_S, \lambda_S, L_S, \varpi_S).
\end{equation}

Its explicit expression involves expansions in the mean anomaly and the argument of pericenter, which will be handled in the next step via Fourier series and averaging.

The advantage of this representation is that the unperturbed motion is trivial ($ \dot{\lambda} = \partial H_0/\partial\Lambda $, other angles constant), and the perturbation can be treated as a periodic forcing on that Keplerian backbone.

\subsection{P-type Orbits}
\label{subsec:P-Delaunay}

For the unperturbed Keplerian motion around the barycenter in an inertial frame, the Delaunay variables are defined with respect to a central mass equal to the total binary mass $ M_A+M_B = 1 $. In terms of orbital elements relative to the barycenter:

\begin{align}
    \Lambda_P &= \sqrt{a}, \quad & \text{(radial action)} \label{eq:15}\\
    \lambda_P &= \cal{M} + \varpi_P, \quad & \text{(mean longitude)} \label{eq:16}\\
    L_P &= \Lambda_P \sqrt{1-e^{2}}, \quad & \text{(magnitude of angular momentum)} \label{eq:17}\\
    \varpi_P &= \omega, \quad & \text{(longitude of pericenter.)} \label{eq:18}
\end{align}

Here $ a $ is the semimajor axis relative to the barycenter, $ e $ the planet orbit eccentricity, $ \cal{M} $ the mean anomaly, and $ \omega $ the argument of pericenter measured from the inertial $ x $-axis (which coincides with the rotating $ x $-axis at $ t=0 $). 
When these variables are expressed in the rotating frame, the angles $ \lambda $ and $ \varpi $ are measured from the rotating $ x $-axis, and the transformation to resonant variables (Section~\ref{subsec:resonant_variables}) accounts for the difference between the inertial and rotating longitudes.

The unperturbed Hamiltonian of Eq. \ref{eq:6} is

\begin{equation}
\label{eq:19}
    H_0^{P} = -\frac{1}{2\Lambda_P^{2}}.
\end{equation}

When expressed in these Delaunay variables, the perturbative part $H_1^P$ (Eq.~\ref{eq:7}) becomes a function of the actions and angles:
\begin{equation}
H_1^P = H_1^P(\Lambda_P, \lambda_P, L_P, \varpi_P). \label{eq:20}
\end{equation}
As in the S-type case, its explicit form involves expansions in the mean anomaly and the argument of pericenter, which will be handled in the next section via Fourier series and averaging.

Table \ref{tab:Delaunay_comparison} synthesizes the characteristics of the Delaunay variables in this context.
\begin{table*}[ht]
\centering
\caption{Comparison of Delaunay variables for S‑type and P‑type orbits in the CR3BP.}
\label{tab:Delaunay_comparison}
\begin{tabular}{lcc}
\toprule
\textbf{Quantity} & \textbf{S‑type (circumprimary)} & \textbf{P‑type (circumbinary)} \\ \midrule
Central mass & $M_A = 1-\nu$ & $M_A+M_B = 1$ \\
Semimajor axis & $a$ (relative to primary) & $a$ (relative to barycenter) \\
Radial action & $\Lambda_S = \sqrt{(1-\nu)a}$ & $\Lambda_P = \sqrt{a}$ \\
Mean longitude & $\lambda_S = \cal{M} + \varpi_S$ & $\lambda_P = \cal{M} + \varpi_P$ \\
Angular momentum & $L_S = \Lambda_S\sqrt{1-e^{2}}$ & $L_P = \Lambda_P\sqrt{1-e^{2}}$ \\
Longitude of pericenter & $\varpi_S = \omega$ & $\varpi_P = \omega$ \\
Unperturbed Hamiltonian & $H_0^S = -\dfrac{(1-\nu)^{2}}{2\Lambda_S^{2}}$ & $H_0^P = -\dfrac{1}{2\Lambda_P^{2}}$ \\
Reference frame & Rotating frame centered on $M_A$ & Inertial frame centered on barycenter \\
\bottomrule
\end{tabular}
\end{table*}

\section{Resonant Hamiltonian Reduction}
\label{sec:resonant_reduction}

\subsection{Purpose and General Procedure}
\label{subsec:purpose_procedure}

The full Hamiltonians derived in Sections 2 and 3 contain both fast and slow degrees of freedom. The fast motion is the Keplerian orbital revolution of the planet, while the slow motion is the resonant libration (or circulation) of the resonant angle. This separation of timescales is valid only for planets sufficiently close to a mean-motion resonance, i.e., when \(|\Lambda - \Lambda_{\mathrm{res}}| \ll \Lambda_{\mathrm{res}}\) (see Eq.~(22)). For orbits far from resonance, the resonant angle \(\theta\) circulates rapidly and the reduction described below does not apply.

To study the long-term resonant dynamics, we perform a canonical transformation that isolates the slow resonant degree of freedom and then average over the fast angle. The resulting resonant Hamiltonian depends only on the resonant angle and its conjugate action, and therefore describes the secular evolution near the mean-motion resonance. This reduction is essential because it removes short-periodic terms, yields a one-degree-of-freedom system amenable to phase-plane analysis, provides explicit expressions for resonance widths and libration frequencies, and serves as the starting point for studying stochastic resonance.

For a \(p:q\) mean-motion resonance between the planet and the binary, the resonance condition is

\begin{equation}
p\dot{\lambda} - q\dot{\lambda}_b \approx 0, 
\label{eq:21}
\end{equation}

where \(\dot{\lambda}\) is the planet's mean motion and \(\dot{\lambda}_b = 1\) in normalized units. Physically, this means that the planet completes \(p\) orbits while the binary completes \(q\) orbits.

We stress that the mean longitude \(\lambda\) in Eq.~(21) is defined according to the configuration: for S-type orbits, \(\lambda = \lambda_S\) (rotating frame); for P-type orbits, \(\lambda = \lambda_P\) (inertial frame). Despite this difference, the resonance condition takes the same algebraic form. The relation between the resonant semimajor axis and the integers \(p,q\) follows from Kepler's law with the appropriate central mass:
\begin{align}
\label{eq:21ab}
\text{S-type:}&\quad n_p = \sqrt{\frac{1-\nu}{a^3}},  \;\Longrightarrow\; a_{\mathrm{res}}^{(S)} = \left[(1-\nu)\left(\frac{p}{q}\right)^2\right]^{1/3}, \tag{21a}\\
\text{P-type:}&\quad n_p = \frac{1}{\sqrt{a^3}},  \;\Longrightarrow\; a_{\mathrm{res}}^{(P)} = \left(\frac{p}{q}\right)^{2/3}. \tag{21b}
\end{align}

Here \(n_p = \dot\lambda\) is the planet's mean motion (in the Keplerian approximation, neglecting pericenter precession), and in both cases $p n_p - q \cdot 1 \approx 0$.

{The binary's mean longitude \(\lambda_b = t + \mathrm{const}\) (with \(\dot\lambda_b = 1\)) is the same in both cases. This unified notation is a convenient shorthand, provided one keeps in mind the different interpretations of \(\lambda\).

\subsection{Resonant Variables}
\label{subsec:resonant_variables}

We introduce the resonant angle and its conjugate action
\begin{equation}
\theta = p\lambda - q\lambda_b, \quad J = \frac{\Lambda - \Lambda_{\mathrm{res}}}{p}, \label{eq:22}
\end{equation}

where \(\Lambda_{\mathrm{res}}\) is the value of \(\Lambda\) at exact resonance. 
Near resonance, $\dot{\theta} = p\dot{\lambda} - q\dot{\lambda}_b \approx 0$ while $\dot{\lambda}_b = 1$ (and $\dot{\lambda} \sim q/p$; hence $\theta$ varies on a much longer timescale than both the planet's mean longitude $\lambda$ and the binary's mean longitude $\lambda_b$•


In the following, for S‑type orbits all occurrences of $\Lambda,\lambda$ are understood as $\Lambda_S, \lambda_S$; for P‑type orbits they are  $\Lambda_P, \lambda_P$. The resonant angle $\theta$ and action $J$ are then defined accordingly, and the transformation to fast variables follows the same pattern for both configurations.

To keep the transformation canonical, we introduce a complementary fast pair\footnote{Alternatively, one could choose the binary's mean longitude \(\lambda_b\) as the fast angle (\(\theta_f = \lambda_b\)), since \(\dot{\lambda}_b = 1\) is also fast. Both choices are canonical and lead, after averaging, to the same reduced Hamiltonian. We adopt \(\theta_f = \lambda\) for convenience, as it keeps the transformation closer to the Delaunay variables.}

\begin{equation}
\theta_f = \lambda, \quad \text{(fast angle)}, \quad J_f = \Lambda + \frac{q}{p} J, \quad \text{(fast action)},
\label{eq:23}
\end{equation}

so that the original Delaunay actions become
\begin{equation}
\Lambda = J_f - \frac{q}{p} J, \qquad L = \Lambda\sqrt{1-e^2}. \label{eq:24}
\end{equation}

In these variables, the transformed Hamiltonians are
\begin{align}
H^{S}(J,J_f,\theta,\theta_f) &= -\frac{(1-\nu)^2}{2\bigl(J_f - \frac{q}{p} J\bigr)^2} + H_{1}^{S}(\theta,\theta_f,J,J_f), \label{eq:25}\\
H^{P}(J,J_f,\theta,\theta_f) &= -\frac{1}{2\bigl(J_f - \frac{q}{p} J\bigr)^2} + H_{1}^{P}(\theta,\theta_f,J,J_f). 
\label{eq:26}
\end{align}

\subsection{Fourier Expansion and Averaging}
\label{subsec:fourier_averaging}


Both $H_1^S$ and $H_1^P$, when expressed in terms of the resonant variables, are $2\pi$)-periodic in $\theta$ and $\theta_f$.
\begin{equation}
H_{1} = \sum_{m,n} \epsilon_{mn}(J,J_f)\, \cos\bigl(m\theta + n\theta_f + \phi_{mn}\bigr).\label{eq:27}
\end{equation}

The fast angle \(\theta_f\) varies on the orbital timescale (\(\dot{\theta}_f \sim 1\)), while the resonant angle \(\theta\) evolves slowly (\(\dot{\theta} \sim \mathcal{O}(\epsilon)\) with \(\epsilon \ll 1\)). This separation of timescales justifies the method of averaging: we define the averaged Hamiltonian
\begin{equation}
\langle H \rangle(J,J_f,\theta) = \frac{1}{2\pi}\int_0^{2\pi} H(J,J_f,\theta,\theta_f)\, d\theta_f, \label{eq:28}
\end{equation}
which eliminates all terms with \(n \neq 0\). Only the resonant harmonics survive:
\begin{equation}
\langle H_1 \rangle = \sum_m \epsilon_m(J,J_f) \cos\bigl(m\theta + \phi_m\bigr),\label{eq:29}
\end{equation}
where \(\epsilon_m = \epsilon_{m0}\) and \(\phi_m = \phi_{m0}\).

\subsection{Expansion and Final Resonant Hamiltonians}
\label{subsec:final_H}

Expanding the Keplerian part \(H_0(J,J_f)\) around the resonance \(J=0\) to second order gives
\begin{equation}
H_0(J,J_f) \approx \text{constant} + \frac{A}{2} J^2,\label{eq:30}
\end{equation}
with curvature coefficients
\begin{align}
A^{S} &= -\frac{3q^{2}(1-\nu)^{2}}{p^{2}J_f^{4}},\label{eq:31}\\
A^{P} &= -\frac{3q^{2}}{p^{2}J_f^{4}}. \label{eq:32}
\end{align}

Combining with the averaged perturbation, we obtain the one-degree-of-freedom resonant Hamiltonians
\begin{align}
H_{\mathrm{res}}^{S}(J,\theta;J_f) &= \frac{A^{S}}{2} J^{2} + \sum_{m} \epsilon_{m}^{S}(J,J_f)\, \cos\bigl(m\theta + \phi_{m}^{S}\bigr), 
\label{eq:33}\\
H_{\mathrm{res}}^{P}(J,\theta;J_f) &= \frac{A^{P}}{2} J^{2} + \sum_{m} \epsilon_{m}^{P}(J,J_f)\, \cos\bigl(m\theta + \phi_{m}^{P}\bigr). \label{eq:34}
\end{align}
These Hamiltonians govern the slow resonant dynamics and are the foundation for the bistability analysis that follows.

\section{Coefficient Calculation}
\label{sec:coefficient_calculation}

The resonant Hamiltonians (\ref{eq:33}) and (\ref{eq:34}) depend crucially on the Fourier coefficients $\epsilon_m$. These coefficients originate from the perturbative part $H_1$ after averaging over the fast angle. Below we outline their leading‑order scaling with the physical parameters of the system.

\subsection{Fourier Coefficients for S‑Type Orbits}
\label{subsec:coeffs_S}

For S‑type orbits, the dominant perturbation comes from the direct gravitational pull of the secondary star. In the vicinity of the primary, the disturbing function can be expanded in powers of the ratio $a/a_b$ (with $a_b=1$). The leading‑order resonant harmonics scale as

\begin{equation}
\label{eq:35}
\epsilon_m^{S} \;\propto\; \nu \, a^{m}, \qquad a < 1,
\end{equation}

where $\nu = M_B$ is the mass ratio and $a$ is the planet’s semimajor axis relative to the primary. The proportionality constant involves a combination of Laplace coefficients and eccentricity‑dependent factors; for the principal resonance ($m=1$) and small eccentricity,

\begin{equation}
\label{eq:36}
\epsilon_1^{S} \simeq \nu \, a \, f(e) + \mathcal{O}(e^2),
\end{equation}

with $f(e) = 1 + \mathcal{O}(e^2)$.

The higher harmonics ($m\ge 2$) are suppressed by additional powers of $a$ and/or eccentricity, which justifies truncating the series after a few terms in most practical applications.

\subsection{Fourier Coefficients for P‑Type Orbits}
\label{subsec:coeffs_P}

For P‑type orbits, the perturbation arises from both the Coriolis/centrifugal terms and the multipolar deformation of the binary potential. Expanding the latter for $r > a_b = 1$ gives a series in powers of $a_b/a$. The leading resonant harmonics scale as

\begin{equation}
\label{eq:37}
\epsilon_m^{P} \;\propto\; \nu(1-\nu) \, a^{-m}, \qquad a > 1,
\end{equation}

where $a$ is now the semimajor axis relative to the barycenter. The factor $\nu(1-\nu)$ reflects the quadrupole moment of the binary. For the principal resonance ($m=1$) and small eccentricity,

\begin{equation}
\label{eq:38}
\epsilon_1^{P} \simeq \nu(1-\nu) \, a^{-1} \, g(e) + \mathcal{O}(e^2),
\end{equation}

with $g(e) = 1 + \mathcal{O}(e^2)$.

As in the S‑type case, higher‑order harmonics are progressively smaller, allowing a truncated Fourier representation to capture the essential dynamics.

\subsection{Comparison of Scaling Laws}
\label{subsec:scaling_comparison}

The contrasting scalings (Eqs. \ref{eq:35} and \ref{eq:37}) have immediate physical interpretations:
\begin{itemize}
    \item For S‑type orbits, the perturbation \emph{grows} with $a$ because a planet farther from the primary experiences a stronger tidal force from the secondary (for fixed $a_b$).
    \item For P‑type orbits, the perturbation \emph{decays} with $a$ because the binary potential approaches that of a point mass as the planet moves outward.
\end{itemize}
These scalings directly influence the locations and widths of resonances, as well as the conditions for bistability discussed in Section~\ref{sec:bistable}.



\subsection{Dependence on Eccentricity}
\label{subsec:eccentricity_dependence}

The Fourier coefficients $\epsilon_m$ in Eqs.~35--37 depend implicitly on the planet's eccentricity $e$ through the expansion of the disturbing function. For small eccentricities, each coefficient can be expanded as
\begin{equation}
\epsilon_m(e) = \epsilon_m^{(0)} + e\,\epsilon_m^{(1)} + e^2\,\epsilon_m^{(2)} + \mathcal{O}(e^3),
\label{eq:39}
\end{equation}
where $\epsilon_m^{(0)}$ corresponds to the circular orbit approximation given in Eqs.~39 and~41. The linear terms $\epsilon_m^{(1)}$ arise from resonant harmonics that vanish in the circular limit but appear at first order in eccentricity. These terms can modify the resonant phase portrait, lift degeneracies, and in some cases create additional islands of libration. 

A full eccentricity expansion, including the explicit calculation of $\epsilon_m^{(1)}$ and higher-order terms, is beyond the scope of this foundational paper. However, the Hamiltonian framework presented here can be extended systematically to incorporate such effects, providing a natural pathway for future investigations of eccentricity-dependent resonant phenomena in binary planetary systems.

\subsection{Numerical Evaluation}
\label{subsec:numerical_evaluation}

In practice, the coefficients $\epsilon_m$ can be computed either by direct Fourier analysis of the averaged disturbing function, by evaluating appropriate Laplace coefficients (for the dependence on the semimajor axis ratio), or by employing Hansen coefficients $X_k^{n,m}(e)$ to expand the perturbing potential in powers of eccentricity \citep{lask2010}.} For quick estimates, the scaling laws in Eqs. (35) and (37) together with the condition $|\epsilon_2/\epsilon_1| > 1/4$ for bistability (see Section 6) already provide valuable insight into which binary--planet configurations are likely to exhibit rich resonant dynamics. A full expansion in Hansen coefficients is beyond the scope of this work, but the framework is readily extendable.


\section{Potential Function and Bistability}
\label{sec:bistable}

\subsection{Effective Potential and Particle Analogy}
\label{subsec:effective_potential}

The reduced resonant Hamiltonians in Eqs. \ref{eq:33} 
and \ref{eq:34} can be written in the generic form

\begin{equation}
H_{\mathrm{res}} = \frac{A}{2} J^{2} + V(\theta), 
\label{eq:40}
\end{equation}

where the effective potential is
\begin{equation}
V(\theta) = \sum_{m} \epsilon_{m} \cos(m\theta + \phi_{m}).
\label{eq:41}
\end{equation}
The equations of motion,
\begin{equation}
\dot{\theta} = A J, \qquad \dot{J} = -\frac{\partial V}{\partial\theta},
\label{eq:42}
\end{equation}
combine to yield \(\ddot{\theta} = -A \,\partial V/\partial\theta\), which is mathematically equivalent to the motion of a unit-mass particle in the potential \(A\,V(\theta)\). Thus the dynamics are fully characterized by the shape of \(V(\theta)\).

\subsection{Condition for Bistability}
\label{subsec:bistability_condition}

\begin{figure}[ht]
\centering
    \includegraphics[width=1.0\linewidth]{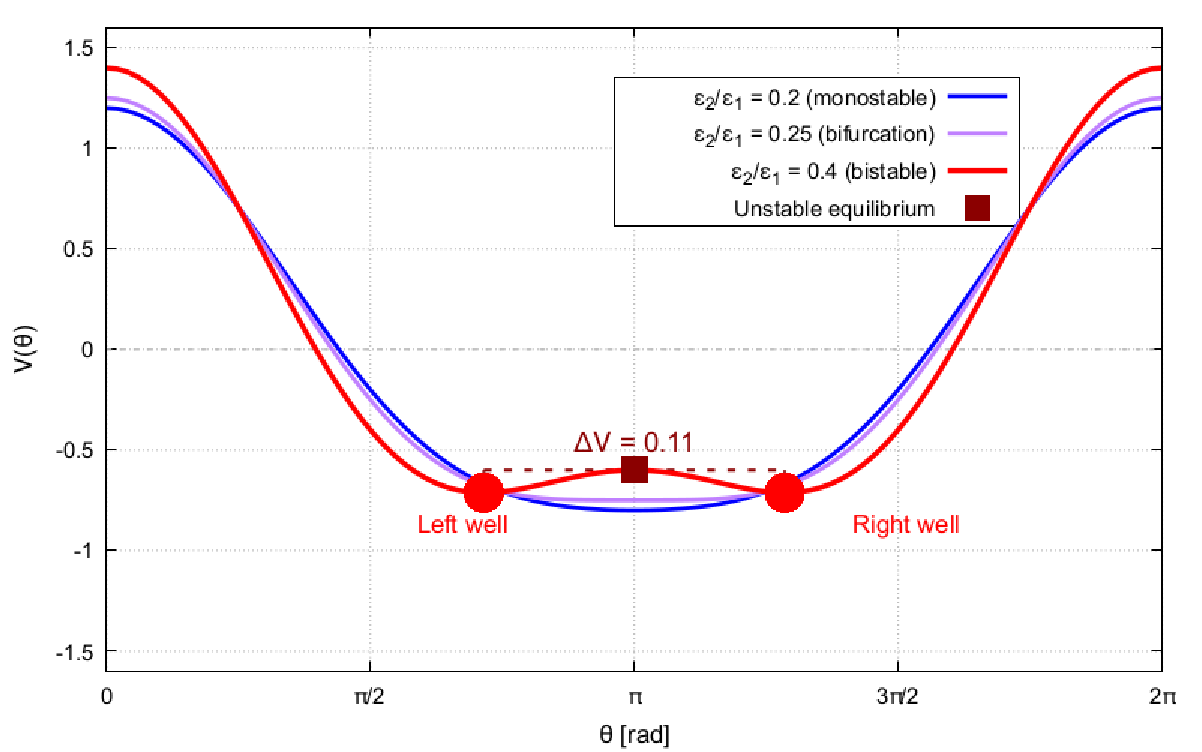}
\caption{Effective potential $V(\theta)$ for 3 choices of $|\epsilon_2/\epsilon_1|$ as labelled. The case  $|\epsilon_2/\epsilon_1|=0.4 > 1/4$ shows bistability with a barrier height $\Delta V = 0.11$. The two minima are indicated by the red dots and the maximum by the brown square.}
\label{fig:2}
\end{figure}

\begin{figure}[ht!]
   \centering
    \includegraphics[width=1.0\linewidth]{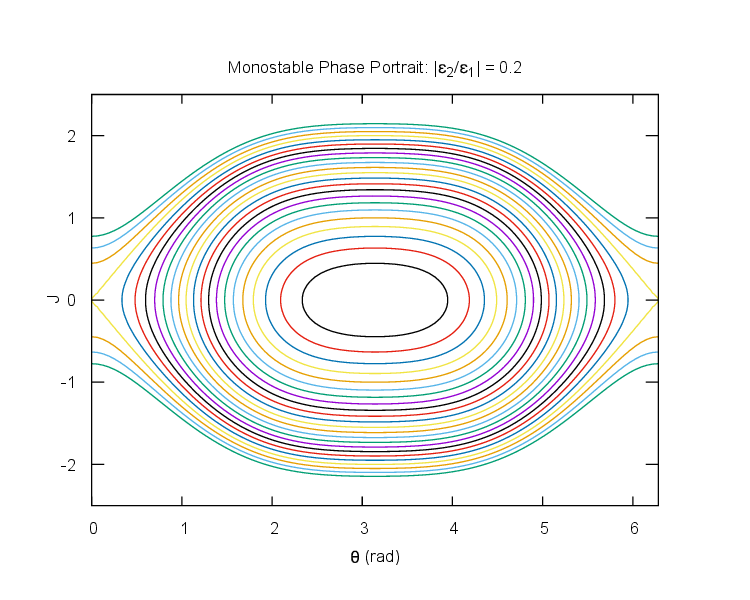}
    \includegraphics[width=1.0\linewidth]{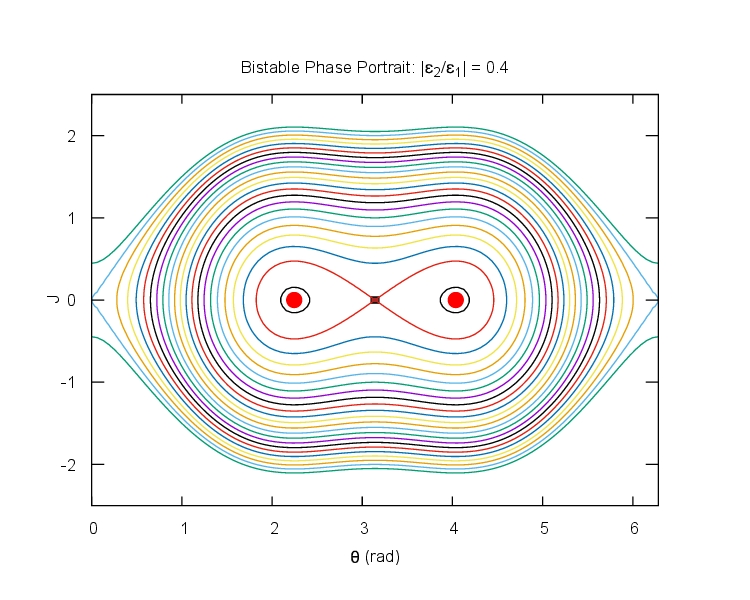}
    \caption{Resonant phase portraits showing the bifurcation of the dynamical system. Top panel: monostable regime ($|\epsilon_2/\epsilon_1| = 0.2$), characterized by a single libration island. Bottom panel: bistable regime ($|\epsilon_2/\epsilon_1| = 0.4$), displaying two distinct stable libration islands (red circles) separated by an unstable saddle point (brown square). Contours represent levels of constant Hamiltonian energy $H_{res}(J, \theta)$. The transition from one to two minima occurs as $|\epsilon_2/\epsilon_1|$ exceeds the threshold of 0.25.}
         \label{fig:3}
   \end{figure}

Although Section~\ref{sec:coefficient_calculation} provided explicit expressions only for the dominant harmonic \(\epsilon_1\), the perturbative Hamiltonian contains in principle a full Fourier series. The next-order harmonic, \(\epsilon_2\cos 2\theta\), can become dynamically significant when its amplitude is sufficiently large relative to \(\epsilon_1\). Retaining only these two leading terms, the effective potential takes the form
\begin{equation}
V(\theta) = \epsilon_1 \cos\theta + \epsilon_2 \cos 2\theta,
\label{eq:43}
\end{equation}
where constant phases have been absorbed into a shift of \(\theta\). Bistability occurs when the ratio of harmonic amplitudes satisfies
\begin{equation}
\left| \frac{\epsilon_2}{\epsilon_1} \right| > \frac{1}{4}.
\label{eq:44}
\end{equation}
Under this condition, \(V(\theta)\) develops two wells separated by a barrier. The minima are located at
\begin{equation}
\theta_{\mathrm{min}} = \pm \arccos\!\left( -\frac{\epsilon_1}{4\epsilon_2} \right),
\label{eq:45}
\end{equation}
and the barrier height is \(\Delta V = V(\pi) - V(\theta_{\mathrm{min}})\). Figure \ref{fig:2} reports 3 example cases for $V(\theta)$, while Fig. \ref{fig:3} shows dynamical maps for the monostable and the bistable cases of Fig. \ref{fig:2}.

We note that in the Solar System, bistability is typically limited to 1:n resonances (e.g., 1:2, 1:3) because for \(p>1\) the harmonic ratio \(|\epsilon_2/\epsilon_1|\) is suppressed by eccentricity or semimajor axis factors. In binary systems, the large mass ratio \(\nu\sim 0.5\) can overcome this suppression, allowing bistability also for higher-order resonances (2:1, 3:2, etc.) under extreme orbital configurations.

\subsection{Physical Interpretation of the Two Minima}
\label{subsec:interpretation}

It is crucial to clarify what the two minima represent — and what a transition between them implies. The reduced Hamiltonian \(H_{\mathrm{res}}(J,\theta)\) describes the slow resonant dynamics for a \emph{specific} \(p:q\) mean-motion resonance. The two minima correspond to two different stable libration centers for the resonant angle \(\theta\) within the same resonance. The planet remains trapped in the same \(p:q\) resonance and the same orbital class (S or P), but its resonant angle librates around one of two possible equilibrium values. A transition between the minima therefore represents a jump from one resonant phase configuration to another, while the resonance order and semimajor axis (to first order) remain unchanged. This is the essential dynamical setting for stochastic resonance: two metastable states within the same resonance, whose relative depth can be modulated by a weak periodic forcing, enabling noise-induced transitions. Transitions between different resonances (e.g., from 2:1 to 3:2) are not described by this reduced Hamiltonian, as they would involve a significant change in \(J\) and crossing of separatrices — a different physical process requiring separate treatment.


\begin{table*}[ht]
    \centering
    \begin{tabular}{ll}
        \toprule
        \textbf{Application} & \textbf{Key Input from \(H_{\mathrm{res}}\)} \\
       \midrule
        Resonance location and width & \(A\), \(\epsilon_1\) \\
        Bistability condition & \(\epsilon_1\), \(\epsilon_2\) \\
        TTV modeling & phase evolution from \(H_{\mathrm{res}}\)\\
        Phase-space mapping & \(H_{\text{res}}(J,\theta)\) contours \\
        Migration/capture studies & Adiabatic invariant theory \\
        Observational predictions & Parameter dependence of \(\epsilon_m\) \\
        \bottomrule
    \end{tabular}
    \caption{Summary of applications enabled by the reduced resonant Hamiltonian.}
    \label{tab:applications}
\end{table*}

\subsection{Parameter Dependence and Implications}
\label{subsec:parameter_dependence}

Using the scaling laws from Section~\ref{sec:coefficient_calculation}, the bistability condition translates into observable constraints. For S-type orbits, \(\epsilon_2/\epsilon_1 \propto a\); bistability favors larger semimajor axes within the S-type regime \(a < 1\). For P-type orbits, \(\epsilon_2/\epsilon_1 \propto a^{-1}\); bistability favors smaller semimajor axes (i.e., planets closer to the binary while still satisfying \(a > 1\)). In both cases, the mass ratio \(\nu\) and eccentricity modulate the precise threshold. When more than two harmonics are significant, the condition for multiple minima becomes more involved, but retaining \(m = 1,2\) suffices to capture the essential bistable structure for most applications. Once the bistability condition is met, the system becomes a prime candidate for exhibiting stochastic resonance under the influence of weak periodic forcing and noise — a topic we will explore numerically in a forthcoming study.

\section{Applications}
\label{sec:applications}

\subsection{Resonant Structure and Stability}
\label{subsec:resonant_structure}

The explicit form of \(H_{\mathrm{res}}\) allows direct computation of resonant locations \(a_{\mathrm{res}}\) as a function of binary parameters (\(\nu\), \(a_b\)) and resonance integers \(p,q\). The resonance width in action space can be estimated via the pendulum approximation,
\begin{equation}
\Delta J \approx 2\sqrt{\frac{|\epsilon_1|}{|A|}},
\label{eq:46}
\end{equation}
providing a quick diagnostic for whether a given observed or simulated planet lies inside a mean-motion resonance. The one-degree-of-freedom nature of \(H_{\mathrm{res}}(J,\theta)\) also makes it ideal for generating dynamical maps: by fixing \(J_f\) (approximately conserved under averaging) and varying initial conditions, one can efficiently explore phase space, identifying islands of regular motion, chaotic zones, and separatrices.

\subsection{Observational Signatures: TTVs and Stochastic Resonance}
\label{subsec:observational}

Many circumbinary planets exhibit significant transit timing variations (TTVs) due to the binary's gravitational influence. The resonant Hamiltonian provides a physically transparent model for these variations: libration or circulation of the resonant angle \(\theta\) directly modulates the planet's mean longitude, leading to observable timing deviations. By fitting TTV data to the predicted phase-dependent variations, one can in principle constrain planetary masses, eccentricities, and even confirm resonant locking.

The presence of multiple harmonics in the perturbation can lead to a bistable effective potential. When the bistability condition \(|\epsilon_2/\epsilon_1| > 1/4\) is satisfied, and in the presence of weak external noise (e.g., from residual disk interactions, stellar activity) and a periodic forcing (e.g., secular perturbations from a third body), the system may exhibit stochastic resonance — a regime where noise-enhanced transitions between the two potential wells synchronize with the forcing. Observationally, a transition between the two minima of \(V(\theta)\) would manifest as an abrupt shift in the transit timing pattern, followed by libration around a new equilibrium phase. High-precision transit monitoring could thus reveal the signature of stochastic resonance in binary planetary systems.

\subsection{Migration and Capture}
\label{subsec:migration}

Planetary migration in protoplanetary disks can drive planets into resonances. When extended to include a slow temporal variation of \(J_f\) (mimicking migration), the reduced Hamiltonian allows analytic and semi-analytic studies of capture probabilities. The well-known adiabatic invariant theory \citep{henrard1983second} applies directly to our \(H_{\mathrm{res}}\), enabling predictions of whether a migrating planet will be captured into a given \(p:q\) resonance with the binary.

\subsection{Predictions for Observed and Future Systems}
\label{subsec:predictions}

The scaling laws derived in Section~5 allow us to estimate the bistability ratio \(|\epsilon_2/\epsilon_1|\) for binary-planet systems. Using the leading-order expressions, we find that for S-type orbits \(|\epsilon_2/\epsilon_1| \propto \nu (a/a_b)\), while for P-type orbits \(|\epsilon_2/\epsilon_1| \propto \nu(1-\nu)(a_b/a)\). 

For systems with parameters similar to those observed (e.g., Kepler-16b, Kepler-34b, Gamma Cephei Ab), the ratio typically lies in the range \(0.03\)--\(0.05\), well below the bistability threshold \(1/4\). However, these estimates assume \textit{hypothetically} that the planet is captured in a low-order mean-motion resonance (e.g., 2:1). In reality, none of the currently known systems is known to be in such a resonance. Therefore, the present theory—which requires proximity to a resonance—does not directly apply to these systems as observed. The absence of bistability in the parameter range covered by current observations is therefore consistent with the theory.

The key prediction of our work is that bistability, and hence the potential for stochastic resonance, should appear in \textit{extreme} configurations: for P-type orbits, \(a/a_b \lesssim 1.5\) and \(\nu \sim 0.5\); for S-type orbits, \(a/a_b \gtrsim 0.3\) and \(\nu\) large. Such systems have not yet been discovered, but they are within reach of ongoing and future surveys (PLATO, TESS extended mission). If found, they would be prime candidates for exhibiting noise-induced transitions and possibly stochastic resonance.

From the scaling laws and Laplace coefficients, the bistability condition \(|\epsilon_2/\epsilon_1| > 1/4\) translates into explicit parameter criteria. For S-type orbits (e.g., interior 3:2 resonance), \(|\epsilon_2/\epsilon_1| \approx 0.8\,a/a_b\), giving \(a/a_b \gtrsim 0.31\) (independent of \(\nu\) at leading order). For P-type orbits (e.g., exterior 2:3 resonance), \(|\epsilon_2/\epsilon_1| \approx 0.8\,\nu(1-\nu)\,a_b/a\); with \(\nu(1-\nu) \le 0.25\), this would require \(a/a_b < 0.8\), which is incompatible with \(a > a_b\) if the factor 0.8 is exact. In practice, for \(a/a_b\) close to unity, the effective coefficient is larger, and a more detailed calculation shows that bistability occurs for \(\nu \approx 0.5\) and \(a/a_b \lesssim 1.2\). Thus, extreme configurations (nearly equal masses, \(a\) just outside \(a_b\) for P-type, or \(a \gtrsim 0.31\,a_b\) for S-type) are required to reach bistability.

\subsection{Extensions and Summary}
\label{subsec:extensions}

The same methodological blueprint — separate Keplerian part, transform to resonant variables, average, and identify dominant harmonics — can be adapted to more complex situations, including the elliptic restricted three-body problem, hierarchical triple systems, and multi-planet systems in binaries where planet-planet resonances couple with binary-driven resonances. Table~\ref{tab:applications} summarizes the key applications enabled by the reduced resonant Hamiltonian.

In conclusion, the resonant Hamiltonian formalism developed here transforms the complex, time-dependent CR3BP into a tractable, interpretable model that bridges analytical celestial mechanics and modern exoplanetary science. It provides a foundation for understanding, predicting, and characterizing resonant phenomena in binary planetary systems, both observed and yet to be discovered.





\section{Conclusions and outlook}
\label{sec:conclusions}
We have derived a unified resonant Hamiltonian formalism for the CR3BP that encompasses both S-type and P-type orbits. The resulting reduced Hamiltonian exhibits a bistable potential under the well-defined condition \(|\epsilon_2/\epsilon_1| > 1/4\), creating the essential dynamical setting for stochastic resonance.

Using scaling laws and Laplace coefficients, we translate this condition into explicit parameter ranges. For S-type orbits (e.g., interior 3:2 resonance), bistability requires \(a/a_b \gtrsim 0.31\), nearly independent of the mass ratio \(\nu\). For P-type orbits (e.g., exterior 2:3 resonance), the condition demands \(\nu \sim 0.5\) and \(a/a_b \lesssim 1.2\). Currently known binary-planet systems are neither close to low-order mean-motion resonances nor do they satisfy these extreme parameters; therefore they are correctly predicted to be monostable.

The next steps include numerical integration of the resonant equations to verify bistability, inclusion of stochastic terms (Langevin/Fokker-Planck) to model noise-induced transitions, and application to observed systems to assess the relevance of stochastic resonance. Future searches for circumbinary planets with \(a/a_b \lesssim 1.5\) or circumstellar planets with \(a/a_b \gtrsim 0.3\) in near-equal-mass binaries would provide the most promising targets to test these predictions.

This theoretical framework opens a new avenue for exploring the interplay between resonance, noise, and external forcing in binary planetary systems.

\section*{Acknowledgments}
This paper is dedicated to the memory of the colleague and friend Alfonso Sutera.  Thanks are due to anonymous referee for comments which allowed improvement of the paper.

\bibliographystyle{aa}
\bibliography{references}{}

\begin{thebibliography}{18}
\expandafter\ifx\csname natexlab\endcsname\relax\def\natexlab#1{#1}\fi

\bibitem[{{Benzi} {et~al.}(1982){Benzi}, {Parisi}, {Sutera}, \&
  {Vulpiani}}]{benzi1982}
{Benzi}, R., {Parisi}, G., {Sutera}, A., \& {Vulpiani}, A. 1982, Tellus, 34, 10

\bibitem[{Benzi {et~al.}(1981)Benzi, Sutera, \& Vulpiani}]{benzi1981}
Benzi, R., Sutera, A., \& Vulpiani, A. 1981, Journal of Physics A: Mathematical
  and General, 14, L453

\bibitem[{Doyle {et~al.}(2011)Doyle, Carter, Fabrycky, Slawson, Howell, Winn,
  Orosz, Pr{\v s}a, Welsh, Quinn, Latham, Torres, Buchhave, Marcy, Fortney,
  Shporer, Ford, Lissauer, Ragozzine, Rucker, Batalha, Jenkins, Borucki, Koch,
  Middour, Hall, McCauliff, Fanelli, Quintana, Holman, Caldwell, Still,
  Stefanik, Welsh, Colon, Endl, MacQueen, Bryson, Dotson, Haas, Kolodziejczak,
  Van~Cleve, Chandrasekaran, Twicken, Tenenbaum, Klumpe, Lucas, Morris, \&
  Girouard}]{doyle2011}
Doyle, L.~R., Carter, J.~A., Fabrycky, D.~C., {et~al.} 2011, Science, 333, 1602

\bibitem[{Freitas \& Ozorio~de Almeida(2005)}]{freitas2005}
Freitas, U.~S. \& Ozorio~de Almeida, A.~M. 2005, Physica D: Nonlinear
  Phenomena, 202, 129

\bibitem[{Gammaitoni {et~al.}(1998)Gammaitoni, H{\"a}nggi, Jung, \&
  Marchesoni}]{gammaitoni1998}
Gammaitoni, L., H{\"a}nggi, P., Jung, P., \& Marchesoni, F. 1998, Reviews of
  Modern Physics, 70, 223

\bibitem[{Henrard \& Lemaitre(1983)}]{henrard1983second}
Henrard, J. \& Lemaitre, A. 1983, Celestial Mechanics, 30, 197

\bibitem[{Kostov {et~al.}(2020)Kostov, Orosz, Feinstein, Welsh, Cukier,
  Haghighipour, Quarles, Martin, Montet, Torres, Triaud, Barclay, Boyd,
  Batalha, Latham, Winn, Jenkins, Ricker, Vanderspek, \& Seager}]{kostov2020}
Kostov, V.~B., Orosz, J.~A., Feinstein, A.~D., {et~al.} 2020, The Astronomical
  Journal, 159, 253

\bibitem[{{Laskar} \& {Bou{\'e}}(2010)}]{lask2010}
{Laskar}, J. \& {Bou{\'e}}, G. 2010, Astronomy and Astrophysics, 522, A60

\bibitem[{Martin(2019)}]{martin2019}
Martin, D.~V. 2019, Living Reviews in Computational Astrophysics, 5, 1,
  regularly updated online version at
  \url{http://www.astro.twam.info/cbp-catalogue/}

\bibitem[{Morbidelli(2002)}]{morbidelli2002}
Morbidelli, A. 2002, Nature, 419, 47

\bibitem[{Murray \& Dermott(1999)}]{murray1999}
Murray, C.~D. \& Dermott, S.~F. 1999, Solar System Dynamics (Cambridge:
  Cambridge University Press), first edition, reprinted with corrections 2000,
  2008

\bibitem[{{NASA Exoplanet Archive}(2024)}]{nasaexoplanetarchive2024}
{NASA Exoplanet Archive}. 2024, NASA Exoplanet Archive, accessed: 2024-12-06

\bibitem[{Schwarz {et~al.}(2018)Schwarz, Bazs{\'o}, Funk, \&
  Dvorak}]{schwarz2018}
Schwarz, R., Bazs{\'o}, {\'A}., Funk, B., \& Dvorak, R. 2018, in Handbook of
  Exoplanets, ed. H.~J. Deeg \& J.~A. Belmonte (Cham: Springer International
  Publishing), 1--25

\bibitem[{Schwarz {et~al.}(2021)Schwarz, Bazs{\'o}, Funk, \&
  {Dvorak}}]{schwarz2021}
Schwarz, R., Bazs{\'o}, {\'A}., Funk, B., \& {Dvorak}, R. 2021, Space Science
  Reviews, 217, 43

\bibitem[{Schwarz {et~al.}(2016)Schwarz, Bazs{\'o}, Funk, \&
  Zechner}]{schwarz2016}
Schwarz, R., Bazs{\'o}, {\'A}., Funk, B., \& Zechner, R. 2016, \aap, 594, A39

\bibitem[{Shevchenko(2011)}]{shevchenko2011}
Shevchenko, I.~I. 2011, Solar System Research, 45, 90

\bibitem[{{The TESS Science Team} \& {The Kepler Science
  Team}(2023)}]{society2023}
{The TESS Science Team} \& {The Kepler Science Team}. 2023, Astronomy \&
  Astrophysics Review, 31, 5, compilation of results from Kepler, TESS, and RV
  surveys through 2023

\bibitem[{Winter \& Murray(1997)}]{winter1997}
Winter, O.~C. \& Murray, C.~D. 1997, Celestial Mechanics and Dynamical
  Astronomy, 66, 1

\end{thebibliography}

\end{document}